\newcommand{\beq}{\begin{equation}}
\newcommand{\eeq}{\end{equation}}
\newcommand{\beqa}{\begin{eqnarray}}

\newcommand{\eeqa}{\end{eqnarray}}

\renewcommand{\beq}{ $$ }
\renewcommand{\eeq}{ $$ }
\renewcommand{\beqa}{\begin{eqnarray*}}
\renewcommand{\eeqa}{\end{eqnarray*}}

\documentstyle[12pt]{article}

\begin{document}
\title{Parallel Implementations of the Split-Step Fourier Method 
for Solving Nonlinear Schr\"odinger Systems}
\author{S.M. Zoldi\\
{\small  zoldi@phy.duke.edu} \\ 
{\small Department of Physics, Duke University}  \\ 
{\small Box 90305, Durham, NC 27708-0305}\\[12pt] 
V. Ruban, A. Zenchuk\\ 
{\small  ruban@itp.ac.ru, zenchuk@itp.ac.ru} \\
{\small L.D.~Landau Institute for Theoretical Physics}\\
{\small Russian Academy of Sciences, 2 Kosygin St., Moscow, Russia}\\[12pt]
S. Burtsev\\ 
{\small  burtsev@cnls.lanl.gov} \\ 
{\small Theoretical Division and Center for Nonlinear Studies}  \\ 
{\small Los Alamos National Laboratory, Los Alamos, NM 87545} \\ 
} 
\date{\today}
\maketitle

\vskip .2cm

\begin{abstract}

We present a parallel version of the well-known Split-Step Fourier method
(SSF) for solving the Nonlinear Schr\"odinger equation, a mathematical
model describing wave packet propagation in fiber optic lines.  The
algorithm is implemented under both distributed and shared memory
programming paradigms on the Silicon Graphics/Cray Research Origin 200.
The 1D Fast-Fourier Transform (FFT) is parallelized by writing the 1D FFT
as a 2D matrix and performing independent 1D sequential FFTs on the rows
and columns of this matrix.  We can attain almost perfect speedup in SSF
for small numbers of processors depending on both problem size and
communication contention.  The parallel algorithm is applicable to other
computational problems constrained by the speed of the 1D FFT.

\end{abstract}

\newpage
\section{Introduction}

The Nonlinear Schr\"odinger equation (NLSE)

\begin{equation}
iA_t + \sigma d \frac{1}{2}A_{xx}+ A^* A^2 = G,
\label{eq:pertNLS}
\end{equation}

is a nonlinear partial differential equation that describes 
wave packet propagation 
in a medium with cubic nonlinearity.  Technologically, the most
important application of NLSE is in the field of nonlinear fiber 
optics~\cite{K_H_book,Agrawal_book}.  The parameter $\sigma$ specifies
the fiber anomalous group velocity dispersion ($\sigma= 1$) or normal
group velocity dispersion ($\sigma= -1$), while the parameter $d$ defines
the normalized absolute value of the fiber's dispersion. The
perturbation $G$ is specified by details of the physical fiber being
studied.

In the special case $G=0$, NLSE is integrable~\cite{Z_S_72} 
and can be solved analytically. In general if 
$G\neq0$ NLSE must be solved numerically.  One of the most
popular numerical methods to solve the perturbed NLSE is the
Split-Step Fourier method (SSF)~\cite{Agrawal_book}.
For small-scale calculations, serial implementations of SSF are
adequate; however, as one includes more physics in the simulation, the
need for large numbers of Fourier modes to accurately solve NLSE 
equation demands parallel implementations of SSF.

Many fiber optics problems demand large-scale numerical simulations
based on the SSF method.  One class of such problems involves accurate
modeling of wave-length division multiplexed
(WDM) transmission systems where many optical channels operate at
their own frequencies and share the same optical fiber.  WDM is
technologically important as it is one of the most effective ways to
increase the transmission capacity of optical 
lines~\cite{K_H_book,Agrawal_book,IIIa}.
To accurately model WDM one needs to
include large numbers of Fourier harmonics in the numerical simulation
to cover the entire transmission frequency band.  Moreover, in WDM
systems different channel pulses propagate at different velocities
and, as a result, collide with each other.  At the pulse collision,
Stokes and anti-Stokes sidebands are generated; these high frequency 
perturbations lead to signal 
deterioration~\cite{Mam_Mol_96a,Ber_Dav_96}.  Another fundamental 
nonlinear effect
called four-wave mixing (FWM)~\cite{Butcher_book}
must be accurately simulated as the FWM components broaden the
frequency domain which requires even larger numbers of Fourier
modes for accurate numerical simulation.

To suppress the FWM~\cite{Mam_Mol_96a,Ber_Dav_96} and make
possible the practical realization of WDM, one can use dispersion
management (concatenation of fiber links with variable dispersion
characteristics).  The dispersion coefficient $d$ in NLSE is now no
longer constant but represents a rapidly varying piecewise constant
function of the distance down the fiber.  As a result, one must take
a small step size along the fiber to resolve dispersion variations and
the corresponding pulse dynamics.  A final reason to include a large 
number of Fourier modes in numerical simulations is to model the
propagation of pseudorandom data streams over large distances.
 
All of the above factors make simulation of NLSE quite CPU
intensive.  Serial versions of the split-step Fourier method in 
the above cases may too be slow even on the fastest modern
workstations.  To address the issue of accurately simulating physical
telecommunication fibers in a reasonable amount of time, we discuss
the parallelization of SSF algorithm for solving NLSE.  Our parallel
SSF algorithm is broadly applicable to many systems and not limited 
to the solution 
of NLSE.  We consider an algorithm appropriate for multiprocessor 
workstations. 
Parallel computing on multiprocessor systems raises
complex issues including solving problems efficiently with 
small numbers of processors, limitations due to the increasingly 
complex memory hierarchy,
and the communication characteristics of shared and distributed 
multiprocessor systems.

Current multiprocessors have evolved towards a generic parallel
machine, which shares characteristics of both shared and distributed memory
computers.  Therefore most commercial multiprocessors support both
shared memory and distributed memory programming paradigms.  The shared
memory paradigm consists of all processors being able to access some amount
of shared data during the program execution.  This addressing of memory on
different nodes in shared memory multiprocessors causes complications in
writing efficient code.  Some of the most destructive complications are:
cache hierarchy inefficiency (alignment and data locality), false sharing of 
data contained in a cache block, and cache thrashing
due to true sharing of data.  Most vendors provide compiler directives to
share data and divide up computation (typically in the form of loop
parallelism) which in conjunction with synchronization directives 
can be used to speed
up many sequential codes.  In distributed memory programming, each
processor works on a piece of the computation independently and must
communicate the results of the computation to the other processors.  
This communication must
be written explicitly into the parallel code, thus requiring more costly
development and debugging time.  The communication is typically handled by
libraries such as the message passing interface (MPI)~\cite{MPI} which
communicates data through Ethernet channels or through the existing memory
system.  Our primary goal is to present a parallel split-step Fourier
algorithm and implement it under these two different parallel programming
paradigms on a 4-processor Silicon Graphics/Cray Research Origin 200
multiprocessor computer.

The remainder of this paper is organized as follows.  In Section 2, 
we recall a few basics of the the split-step Fourier method.  In Section 3, we
introduce the parallel algorithm for SSF.  Timing results and
conclusions are given in Sections 4 and 5, respectively.

\section{ Split-Step Fourier Method}

The Split-Step Fourier (SSF) method is commonly used to
integrate many types of nonlinear partial differential equations.  In
simulating Nonlinear Schr\"odinger systems (NLS) SSF is predominantly
used, rather than finite differences, as SSF is often more
efficient~\cite{Taha}.  We remind the reader
of the general structure of the numerical
algorithm~\cite{Agrawal_book}.

NLS can be written in the form:
$$
\frac{\partial A}{\partial t}=(L+N)A,
$$
where $L$ and $N$ are linear and nonlinear parts of the equation.  The
solution over a short time interval $\tau$ can be written in the form
$$
A(t+\tau,x)=\exp(\tau L) \exp(\tau N)A(t,x)
$$
where the linear operator in NLS acting on a spatial field~$B(t,x)$ 
is written in Fourier space as,
$$
\exp(\tau L)B(t,x)=F^{-1}\exp(-ik^2\tau) F B(t,x)
$$
where $F$ denotes Fourier transform (FT), $F^{-1}$ denotes the inverse Fourier
transform, and $k$ is the spatial frequency.

We split the computation of $A$ over time interval $\tau$ into 4
steps:

\vspace{0.5cm}

{\bf Step 1.} {\it Nonlinear step:} Compute $A_1=\exp(\tau N)A(t,x)$
(by finite differences);

{\bf Step 2.} {\it Forward FT:} Perform the forward FFT on
$A_1$: $A_2=F A_1$;

{\bf Step 3.} {\it Linear step:} Compute $A_3= \exp(\tau L) A_2$;
  
{\bf Step 4.} {\it Backward FT:} Perform the backward FFT on
$A_3$: $A(t+\tau)=F^{-1} A_3$;

\vspace{0.5cm}

To discretize the numerical approximation of the above algorithm, the
potential $A$ is discretized in the form: $ A_l=A(lh);l=0,\dots,N-1,
$ where $h$ is the space-step and $N$ is the total number of spatial
mesh points.

The above algorithm of the Split-Step Fourier (SSF)
method is the same for both sequential and parallel code.  Parallel 
implementation of this algorithm involves parallelizing each 
of the above four steps.

\vskip 0.5cm
\section{ The Parallel Version of the Split-Step Fourier (SSF) Method}
\vskip 0.5cm

By distributing computational work between several processors, one can
often speed up many types of numerical simulations. A major prerequisite in
parallel numerical algorithms is that sufficient independent computation be
identified for each processor and that only small amounts of data are communicated between
periods of independent computation.  This can often be done trivially
through loop-level parallelism (shared memory implementations) or
non-trivially by storing true independent data in each processor's local
memory.  For example, the nonlinear transformation in the SSF algorithm 
involves the independent computation over subarrays of spatial
elements of~$A(l)$. Therefore $P$ processors each will
work on sub-arrays of the field $A$, e.g., the first processor
updates $A_{0}$ to $A_{(N/P-1)}$, the second processor updates $A_{N/p}$ to
$A_{2(N/p)-1}$, etc.

In the 1D-FFT, elements of $(FA)_k$ can not be computed in a
straightforward parallel manner, because all elements $A_l$ are used
to construct any element of $(FA)_k$. The problem of 1D FFT
parallelization has been of great interest for 
vector~\cite{Swartztrauber_87,Aver_90} and distributed memory
computers~\cite{Dubey_94}.  These algorithms are highly architecture
dependent, involving efficient methods to
do the data re-arrangement and transposition phases of the 1D FFT.
Communication issues are paramount in 1D FFT parallelization and in
the past have exploited classic butterfly communication patterns to
lessen communication costs~\cite{Dubey_94}.  However, due to a rapid
change in parallel architectures, towards multiprocessor
systems with highly complex memory hierarchies and communication
characteristics, these algorithms are not directly applicable to many
current multiprocessor systems.  Shared memory multiprocessors often
have efficient communication speeds, and we therefore implement the
parallel 1D FFT by writing~$A_{l}$ as a two dimensional array, in which 
we can identify independent serial 1D FFTs of rows and columns of this
matrix.  The rows and columns of the matrix~$A$ can be
distributed to divide up the computation among several processors.  Due to
efficient communication speeds, independent computation stages, and
the lack of the transposition stage of the 1D FFT in SSF computations,
we show that this method exploits enough independent computation to result
in a significant speedup using a small number of processors.

\subsection{Algorithm of Parallel SSF} 

The difficulty parallelizing the split-step Fourier algorithm 
is in steps {\bf 2} and {\bf 4}, 
as the other two steps can be trivially evolved due to the natural
independence of the data $A$ and $A_2$.  In {\bf Step 2} and {\bf Step 4}
there are non-trivial data dependences over the entire range $0<=l<=N$ of
$A_{1}(l)$ and $A_{3}(l)$ which involve forward and backward Fourier 
Transforms (FFT and BFT).  The discrete forward Fast Fourier Transform (FFT)
is of the form
$$
F(k)=\sum_{l=0}^{N-1}A(l)\exp\left(-\frac{2\pi i l k}{N}\right)
$$
which requires all elements of $A(l)$.  Several researchers have
suggested parallel 1D Fast Fourier Transform
algorithms~\cite{Swartztrauber_87,Aver_90,Dubey_94}, but to date
there exist no vendor-optimized parallel 1D FFT algorithms.
Therefore implementations of these algorithms are highly 
architecture dependent.  Parallel 1D FFT algorithms must deal with
serious memory hierarchy and communication issues in order to achieve
good speedup.  This may be the reason why vendors have been slow to
support the computational community with fast parallel 1D FFT
algorithms. However, we show below that we can get significant
parallel speedup due to the elimination of the
transposition stage in 1D FFT for SSF methods and due to exploitation of
independent computation by performing many sequential 
1D FFTs on small subarrays of~$A(l)$.

Our method of parallelizing the SSF algorithm 
requires dividing the 1D array $A(l)$ into subarrays 
which are processed using vendor optimized {\em sequential}
1D FFT routines.  We assume the 1D array $A$ is of the dimension~$N$ of 
the product of two integer numbers: $N=M_0 \times M_1$.  Therefore
$A$ can be written as a 2D matrix of size~$M_0 \times M_1$.
As a result, we can reduce the expression for the
Fourier transform of the array $A$ to the form 
\begin{equation}
F(M_1 k_1+k_0)=\sum_{n_0=0}^{M_0-1}\sum_{n_1=0}^{M_1-1} A(M_0n_1+n_0)
exp\left ( \frac{-2\pi i}{M_0M_1}(M_0n_1+n_0)(M_1 k_1+k_0) \right )
\eeq
\beq
=\sum_{n_0=0}^{M_0-1} f(k_0,n_0)exp\left(\frac{-2\pi i}{M_0M_1}n_0k_0\right)
exp\left(\frac{-2\pi i}{M_0}n_0k_1\right)
\label{eq:re}
\end{equation}
where $F$ is the Fourier transform of $A$ and $f$ is the result of $M_1$- size
Fourier transform of $A(M_0n_1+n_0)$ with fixed $n_0$
\begin{equation}
f(k_0,n_0)=\sum_{n_1=0}^{M_1-1} A(M_0n_1+n_0)
 exp\left(\frac{-2\pi i}{M_1}n_1k_0\right)
\end{equation}
\begin{equation}
n_0,k_1=0,...,M_0-1 \hspace{15mm}n_1,k_0=0,...,M_1-1. 
\end{equation}
The reduced expression~Eq.~(2) demonstrates that 
the $N=M_0 * M_1$ Fourier transform 
$F$ is obtained by making $M_0$ size Fourier transforms of 
$f(k_0,n_0)exp\left(\frac{-2\pi i}{N}n_0k_0\right)$ for fixed $k_0$. 
Therefore the 1D array $A$ is written as a 2D matrix $a_{jk}$ 
of size $M_0 \times M_1$ with elements $(A(0),...,A(M_0-1))$ in the 
first column, $(A(M_0),...,A(2M_0-1))$ in the second column, 
etc. We use this matrix $a_{jk}$ in our parallel FFT-algorithm:

\vspace{0.5cm}

{\bf Step 1.} Independent $M_1$-size FFTs on rows of~$a_{jk}$.

{\bf Step 2.} Multiply elements $a(j,k)$ by a factor 
$E_{jk}=exp(-(2\pi i /N)\cdot j \cdot k)$

{\bf Step 3.} Independent $M_0$-size FFTs on columns of~$a_{jk}$.

\vspace{0.5cm}

The result of {\bf Step 1 - Step 3} is the $N=M_0*M_1$ 1D Fourier transform 
of $A$ stored in {\it rows}: $(F(0),...,F(M_1-1))$ in the first row, 
$(F(M_1),...,F(2M_1-1))$ in the second row, and so on.  To regain
the proper ordering of $A$ (how elements were originally stored
in matrix $a_{jk}$) requires a transposition of the matrix which
is the last step in a parallel FFT algorithm.

In the SSF method, the transposition is {\em not 
necessary} as we apply a linear operator~$L(k)$ and then take the 
steps: {\bf Step 1 - Step 3} in reverse order. 
This avoids the transposition because one can
define a transposed linear operator array and 
multiply $a_{jk}$ by this transposed linear operator. 
Then {\bf Step 1 - Step 3} are performed 
in reverse order with the conjugate of the exponential term in
{\bf Step 2}.

\vspace{0.5cm}

The complete SSF parallel algorithm consists of the following steps:
\vspace{0.5cm}

{\bf Step 1.} nonlinear step

{\bf Step 2.} row-FFT

{\bf Step 3.} multiply by $E$

{\bf Step 4.} column-FFT

{\bf Step 5.} linear step (transposed linear operator)

{\bf Step 6.} column-BFT

{\bf Step 7.} multiply by $E^{*}$ (the complex conjugate of $E$) 

{\bf Step 8.} row-BFT

\vspace{0.5cm}

The parallelization is due to the natural independence of operations
in steps 1, 3, 5, and 7 and by the row and column subarray FFTs in
steps 2, 4, 6, and 8.  The row and column subarray FFTs of size $M_1$
and $M_0$ are performed independently with serial optimized 1D FFT
routines.  Working with subarray data, many processors can be used to
divide up the computation work resulting in significant speedup if 
communication between processors is efficient.  Further, smaller
subarrays allows for better data locality in the primary and secondary
caches.  The implementation details of the shared-memory and the
distributed memory parallel SSF algorithm outlined above depend on
writing {\bf Steps 1 - 8} using either shared memory directives or
distributed memory communication library calls (MPI).

\subsection{Shared Memory Approach}

Much of the SSF parallel algorithm outlined above can be implemented
with ``\$doacross'' directives to distribute independent loop
iterations over many processors.  The FFTs of size $M_0~{\rm and}~M_1$
are implemented by distributing the 1D subarray FFTs of rows and
columns over the P available processors.  The performance can be improved
drastically by keeping the same rows and columns local in a
processor's secondary cache to alleviate true sharing of 
data from dynamic assignments of sub-array FFTs 
by the ``\$doacross'' directive.
The subarray FFTs are performed using vendor optimized sequential 
1D FFT routines which are designed specifically for the architecture.

It is efficient to perform all column operations ({\bf Steps 3 - 7})
in one pass: copying a column into local sub-array $S$ contained
in the processor's cache and
in order, multiply by the exponents in {\bf Step 3}, perform the
$M_0$-size FFT of $S$, multiply by the transposed linear operator
$\exp (\tau L)$, invert the $M_0$-size FFT, multiply by the conjugate
exponents, and finally store $S$ back into the same column of $a$.  
This allows for efficient use of the cache, reducing false/true sharing 
as we perform many operations on each subarray.

\subsection{Distributed Memory Approach}

The Massage Passing Interface (MPI) is a tool for distributed 
parallel computing which has become a standard used on a variety
of high-end parallel computers to weakly coupled distributed networks
of workstations (NOW)~\cite{MPI}.  In distributed parallel programming, 
different processors work on completely independent data and 
explicitly use send and receive library calls to communicate 
data between processors.  

To implement the distributed parallel SSF algorithm for the Nonlinear 
Schr\"odinger system (NLS), one needs to distribute the rows
of array $A$ among all P available processors.  Then 
{\bf Steps 1 - 3} can be executed without communication 
between processors.  
After these steps, it is necessary to endure the communication cost 
of redistributing the elements of $A$ among the P processors.  Each 
processor must send a fraction ${\frac{1}{P}}$ of its data to each of the other 
processors.  Then each processor will have the
correct data for {\bf Steps 4 -7} and column operations are performed 
independently on all P processors.  Finally, there is
a second redistribution prior to  {\bf Step 8}.  
To make $T$ steps of the SSF algorithm, we use
the following scheme:
\vspace{0.5cm}

\centerline{{\bf subroutine Distributed SSF}}

{\it distribute rows among processors}

{\bf Step 1.} nonlinear step

{\bf Step 2.} row-FFT

{\bf Step 3.} multiply by a factor $E$

\vspace{0.5cm}

{\bf for $i=1$ to $T-1$ do}

\hspace{1cm}{\it data redistribution}

\hspace{1cm}{\bf Step 4.} column-FFT

\hspace{1cm}{\bf Step 5.} linear step

\hspace{1cm}{\bf Step 6.} column-BFT

\hspace{1cm}{\bf Step 7.} multiply by a factor ${E}^\ast$ 

\hspace{1cm}{\it data redistribution}

\hspace{1cm}{\bf Step 8.} row-{BFT}

\hspace{1cm}{\bf Step 1.} nonlinear step

\hspace{1cm}{\bf Step 2.} row-FFT

\hspace{1cm}{\bf Step 3.} multiply by a factor $E$

{\bf end do}

\vspace{0.5cm}

{\it data redistribution}

{\bf Step 4.} column-FFT

{\bf Step 5.} linear step

{\bf Step 6.} column-BFT

{\bf Step 7.} multiply by a factor ${E}^\ast$ 

{\bf Step 8.} row-{BFT}

\centerline{{\bf end}}

\vspace{0.25in}

The large performance cost in this algorithm is the redistribution of 
data between row and column operations.  If the row and
column computational stages result in significant speedup compared
to the communication expense of redistributing the matrix data, then 
this algorithm will be successful.  This depends 
{\bf crucially} on fast communication between processors which is
usually the case for shared memory multiprocessors and less
so for NOW computers.

\section{Results}

We performed timings of the parallel SSF algorithm on the Silicon
Graphics/Cray Research Origin 200 multiprocessor.  The Origin 200 was
used because it allows for both shared and distributed memory parallel
programming and models a generic multiprocessor.  The Origin
200 is efficient at fine-grained parallelism which typically makes
shared memory programming both efficient and easy.  The Origin 200
workstation used in this study consisted of four MIPS R10000 64-bit
processors (chip revision 2.6) with MIPS R10010 (revision 0.0)
floating point units running at 180MZ.  The primary cache consisted of
a 32KB 2-way set-associative instruction cache and a 32KB 2-way
set-associative data cache.  Each processor also had a 1MB 2-way
set-associative secondary cache.  The machine had a sustained 1.26GB/sec 
memory bandwidth and 256MB of RAM. 

The operating system was IRIX 6.4.  We used a Mongoose f77 version
7.10 Fortran compiler.  For the parallel programming we used MPI
version 1.1 for the distributed computing and the native
``\$doacross'' and synchronization directives provided by the f77
compiler for shared memory programming.  All timings are of the total
wall-clock time for the code to both initialize and execute.

\subsection{Timings}

For the following timings, we use~$M_0=M_1=2^K$, so that the entire 1D
array is of size $N=2^{2K}$.  The one-processor implementation of parallel
SSF was 10\% to 20\% faster than serial SSF code using
vendor optimized 1D FFTs of the entire array of size~$N=2^{2K}$.  This
improvement is due to better cache coherence using smaller subarrays, as an
entire subarray can be contained in the L1 cache and is due to the fact
that the single processor parallel SSF does not do the transposition stage
of the 1D FFT.  All timings are compared to the {\em one-processor parallel
code} at the same optimization level (compared to sequential SSF the below
speedups are even more impressive).  For shared memory parallel
implementations, we find over the range of~$2^{12} < N < 2^{18}$ that two
node SSF implementations have good speedup (SU) with a maximum speedup
at~$N=2^{16}$.  Using four nodes, for small array sizes we have $1/4$ less
work per processor, but more contention due to the sharing of pieces of
data contained in the secondary caches of four different processors.  At
$N=2^{16}$, we again see the maximum speedup (now for 4 nodes), 
reflecting that the ratio of
computational speed gain to communication contention is 
optimal at this problem size.

\vspace{0.5cm}

{\bf Shared Memory}
$$
\begin{array}{|c|c|c|c|c|}
\mbox{array size ($N$)}&N=2^{12}&N=2^{14}&N=2^{16}&N=2^{18}\cr
\mbox{number of steps ($T$)}&
T=8000 & T=2000 & T=500 & T=125\cr
{T_{1pr.}}~({\rm sec})&
49.5&51.5&65.5&97.5\cr
{T_{2pr}}~({\rm sec})&
29.5&30.5&33.5&61.0\cr
{T_{4pr}}~({\rm sec})&
19.5&18.5&19.5&34.5\cr
{\rm SU}={T_{2pr}/T_{1pr.}}&
1.7&1.7&2.0&1.6\cr
{\rm SU}={T_{4pr}/T_{1pr.}}&
2.5&2.8&3.4&2.8
\end{array}
$$

\vspace{0.5cm}

Under the shared memory programming model, subarrays are continually
distributed among processors to divide up the computational work.  Data
in a single subarray may be contained on one or more processors requiring
constant communication.  The data contained in each processor's L2
cache is of size~O(N/P), where P is the number of processors.  Contention
in the memory system is modeled as being proportional to $O((N/P)^{2})$ 
which reflects the communication congestion 
for sharing data of large working sets.  Further unlike the serial code, the parallel
code endures a communication time to send data between processors 
proportional to~$O(N/P)\tau_c$, where~$\tau_c$ is the time to transfer
a L2 cache block between processors.  Finally, the time
to perform the 1D FFT is approximately $N Log(N) \tau_M$, where~$\tau_M$ 
is the time to perform a floating point operation.  A simple formula for 
the speedup (SU) of the shared memory FFT is
\begin{equation}
SU = {\tau_M N Log(N) \over \tau_M N Log(\sqrt{N})/P + 
\tau_c N/P + f (N/P)^2)},
\end{equation}
where $f$ is a small number reflecting contention in the
communication system.  If $N=2^{K}$ we can simplify the above expression, 
\begin{equation}
SU = {2P \over (1 + \xi /K + f 2^{K}/(PK))},
\end{equation}
where the constants are absorbed into $f$ and $\xi$.
With $f=0$ 
(no contention) one predicts for fixed P that the speedup increases for 
larger and larger problem size~$N$.  However, for~$f \ne 0$ 
the speedup eventually decreases with larger~$N$ due to contention 
of communicating small pieces
of subarray data between arbitrary processors.  This equation reflects the
trend seen in our empirical data of speedup for shared memory SSF, where 
speedup attains a maximum with problem size at $N=2^{16}$.

The above SU formula must be reinterpreted for distributed SSF due to
the implicit independent computational stages where no data is
communicated between processors unlike shared memory SSF.  Distributed
SSF uses communication stages to send data between processors and does
not involve contention due to sharing data between
P processors during computation stages.
Distributed MPI timings are compared to a {\em one-processor
MPI code} at the same optimization level.  The MPI one-processor code
was faster than one-processor shared memory code, as it did not have
synchronization steps.  The parallel timings were typically
faster than the shared memory parallel code, except for the
$N=2^{16}$ array size for which the shared memory code did slightly
better.  We find that for distributed memory parallel
implementations of SSF over the range of~$2^{12} < N < 2^{18}$ 
two-node implementations have good speedup with maximum
speedup at~$N=2^{14}$, beyond which the communication cost
increases and the computation/communication ratio decreases 
for larger problem size.  The communication cost is different 
in the
MPI case than for shared memory, as data is communicated
in ``communication stages'' so less than perfect speedup (SU) is
due to the volume of data communicated between processors 
in redistribution stages.
Using four nodes, we find that the speedup increases with the 
working set~$N$.  This 
is due to both making the computation stages faster~O(NLog(N)/8) 
and by communicating only O(N/16) of data between single processors in 
the redistribution stage.  For small problem size there is 
not enough work to make dividing the problem among 4 processors
beneficial.  The speedup in 
the distributed SSF algorithm is attributed to the independence of 
data contained in a processor's local cache between data 
rearrangement stages, which is not true for the dynamic assignment
and sharing of subarray data throughout computational stages in
shared memory SSF implementations.

\vspace{0.5cm}

{\bf Distributed Memory (MPI)}
$$
\begin{array}{|c|c|c|c|c|}
\mbox{array size ($N$) and}&N=2^{12}&N=2^{14}&N=2^{16}&N=2^{18}\cr
\mbox{number of steps ($S$)}&
S=8000 & S=2000 & S=500 & S=125\cr
{T_{1pr.}}~({\rm sec})&
37.9&44.5&59.4&92.4\cr
{T_{2pr}}~({\rm sec})&
24.7&25.4&34.9&65.9\cr
{T_{4pr}}~({\rm sec})&
18.8&16.3&20.1&26.8\cr
{{\rm SU}=T_{2pr}/T_{1pr.}}&
1.5&1.8&1.7&1.4 \cr
{{\rm SU}=T_{4pr}/T_{1pr.}}&
2.0&2.7&3.0&3.4
\end{array}
$$

\vspace{0.5cm}

These results are encouraging in that the speedup in multiprocessor SSF
implementations is considerable.
Speedup over sequential code using vendor optimized full array 1D FFT
is even greater.  We recommend implementing the parallel SSF algorithm
even on sequential machines due to the 10\% to 20\% speedup over
optimized 1D sequential SSF algorithms.  This reflects a
better use of the L1 cache and data locality by using small subarrays
and removing the transposition stage of the 1D FFT in SSF.  For shared
memory implementations of parallel SSF, the maximum speedup requires
balancing contention in the communicating data contained over more
than one processor to the computation performance gain of using small
subarrays. For the distributed parallel SSF there is more data
locality as data is distributed statically prior to the computational
stages.  This division of computation and communication
stages is different than for shared memory SSF which dynamically
distributes subarray FFTs and shares data on more than one processor.
Distributed SSF speedup is a function of the number of processors~$P$
which reduces both the computational time and communication 
volume between single processors.  The speedup of the parallel SSF is 
strongly dependent on reducing communication time and contention 
in the multiprocessor.

\section{ Conclusions}

Multiprocessor systems occupy the middle ground in computing between
sequential and massively parallel computation.  In multiprocessor computing,
one wants to write code to take advantage of between 2 and 16 processors to
get good speedups over sequential code.  Our parallel SSF method is
designed for small numbers of tightly integrated processors to divide the
1D FFT into many subarray FFTs performed on P processors.  The speedup
depends on optimizing the computational speed gain to communication cost
in order to speedup traditionally sequential numerical code.  The shared
memory parallel SSF algorithm does not scale with problem size~$N$ as
subarray data is distributed over more than one processor causing increases
in contention due to gathering large amounts of subarray data from many
processors.  The distributed memory parallel SSF algorithm
uses independent data during computational stages and then uses
expensive data redistribution stages.  The communication cost of the data
redistribution stages can be reduced by using more processors, which also
decreases the time for the computation stage.  Our results suggest that
nearly perfect speedup can be achieved over sequential SSF algorithms by
tuning the number of processors and problem size.  The significant speedup
over sequential code is broadly applicable to many sorts of code which
depend crucially on speeding up the sequential 1D FFT and should be
explored for other numerical algorithms.
  
\vskip 0.3cm
{\bf ACKNOWLEDGMENTS} 
 
This work has  been 
performed under the auspices of the US Department of 
Energy under contract W-7405-ENG-36 and the Applied Mathematical 
Sciences contract KC-07-01-01.

V. Ruban and A. Zenchuk   thank the
Theoretical Division Group T-7 and Center for Nonlinear Studies at 
the Los Alamos National Laboratory for their hospitality
and financial support during the summer of 1997.  S.M Zoldi was
supported by the Computational Science Fellowship Program of the
Office of Scientific Computing in the Department of Energy.  S.M Zoldi
thanks the Center of Nonlinear Studies for their hospitality during 
the summer of 1997.  Useful conversations with Richard Peltz 
are acknowledged.

%
%
%
%


\newpage

\begin{thebibliography}{12}

\bibitem{K_H_book}
A. Hasegawa and Y. Kodama, {\it Solitons in Optical Communication}, 
Oxford Univ. Press (1995).

\bibitem{Agrawal_book}
G.P. Agrawal, {\it Nonlinear Fiber Optics}, Acad. Press, (1994) 2nd edition.

\bibitem{Z_S_72} 
V.E. Zakharov and A.B. Shabat,  
{\it Sov. Phys. JETP}, {\bf 34}, 62 (1972).

\bibitem{IIIa}
{\it ``Optical Communications IIIa''}, ed. by I.P. Kaminow and T.L. Koch,
Acad. Press, (1997). 

\bibitem{Mam_Mol_96a}
P.V.  Mamyshev and L.F.  Mollenauer, {\it Opt. Lett. }{\bf 21}, (6), 
396 (1996).

\bibitem{Ber_Dav_96}
N.S. Bergano and C.R. Davidson, {\it J. of Lightwave Technology},
{\bf 14}, (6), 1299 (1996).

\bibitem{Butcher_book}
P.N. Butcher and D. Cotter, {\it The Elements of Nonlinear Optics},
Cambridge Studies in Modern Optics, v.9, (1993).

\bibitem{MPI}
M. Snir, S. Otto, S.H Lederman, D. Walker, J. Dongarra,
{\it MPI: The Complete Reference}, MIT Press, London, (1996).

\bibitem{Taha}
T.R Taha and M.J. Ablowitz,
{\it J. Comp Physics}, {\bf 55}, 203 (1984).

\bibitem{Swartztrauber_87}
P.N Swartztrauber,
{\it Parallel Computing}, {\bf 5}, 197 (1987).

\bibitem{Aver_90}
A. Averbuch, E. Gabber, B. Gordissky, Y. Medan,
{\it Parallel Computing}, {\bf 15}, 61 (1990).

\bibitem{Dubey_94}
A. Dubey, M. Zubair, C.E. Grosch,
{\it Parallel Computing}, {\bf 20}, 1697 (1994).

\end{thebibliography}
\end{document}